\documentclass[aps, prd, amsmath,amssymb,notitlepage, twocolumn, nofootinbib]{revtex4-1}

\usepackage{subfigure}
\usepackage{amssymb}
\usepackage{graphicx}
\usepackage{color}
\usepackage{amsmath}
\usepackage[colorlinks=true,linkcolor=blue,citecolor=blue,urlcolor=blue]{hyperref}
\usepackage{framed}
\usepackage{gensymb}
\usepackage{comment}

\newcommand{\be}{\begin{equation}}
\newcommand{\ee}{\end{equation}}

\begin{document}

\title{Nonlocal Control of Dissipation with Entangled Photons}

\author{Charles Altuzarra $^{1,2}$, Stefano Vezzoli$^{1}$, Joao Valente$^{3}$,  
	Weibo Gao$^{1}$, Cesare Soci$^{1}$, Daniele Faccio$^{4}$, Christophe Couteau$^{1,2,5,*}$}

\affiliation{ 1 Centre for Disruptive Photonic Technologies, TPI, Nanyang Technological University, 637371, Singapore\\
	2 CINTRA, CNRS-NTU-Thales, CNRS UMI 3288, Singapore\\
	3 Optoelectronics Research Centre and Centre for Photonic Metamaterials, 
University of Southampton, Southampton, UK\\
	4 Institute for Photonics and Quantum Sciences and SUPA, Heriot-Watt University, Edinburgh EH14 4AS, UK \\
	5 Laboratory for Nanotechnology, Instrumentation and Optics, ICD CNRS UMR 6281, 
University of Technology of Troyes, Troyes, France\\
	* Corresponding author: christophe.couteau@utt.fr}

\begin{abstract}
	
Quantum nonlocality, i.e. the presence of strong correlations in spatially separated systems which are forbidden by local realism, lies at the heart of quantum communications and quantum computing. Here, we use polarization-entangled photon pairs to demonstrate a nonlocal control of absorption of light in a plasmonic structure. Through the detection of one photon with a polarization-sensitive device, we can almost deterministically prevent or allow absorption of a second, remotely located photon. We demonstrate this with pairs of photons, one of which is absorbed by coupling into a plasmon of a thin metamaterial absorber in the path of a standing wave of an interferometer. Thus, energy dissipation of specific polarization states on a heat-sink is remotely controlled, promising opportunities for probabilistic quantum gating and controlling plasmon-photon conversion and entanglement.

\end{abstract}

\date{\today}
\maketitle 

{\textit{Introduction.---}} One of the quintessential aspects of quantum mechanics is the existence of entangled states whereby the classical description of a particle being in a well-defined state is replaced with a quantum description based on superposition of states. Moreover, quantum entanglement provides a unique route for nonlocal correlations between remote particles such as photons. Beyond the relevance to the fundamental questions of quantum physics \cite{1,2,3,4}, nonlocality is a resource for a number of applications such as quantum teleportation, quantum erasure, and interaction-free measurements \cite{5,6,7,8,9,10}. In addition to demonstrating a new application of nonlocality, our work presented here is also relevant to the field of quantum plasmonics. This rather recent field emerged when Altewischer \emph{et al.} \cite{11} showed that light passing through a metallic nanohole array conserved the quantum state of entangled photons. In parallel, Lukin \emph{et al.} demonstrated that single plasmons can be generated from single photons as a process that can be deterministic \cite{12,13}. More recently, Hong-Ou-Mandel two-photon quantum interference experiments were carried out with plasmons \cite{14,15}, thereby providing experimental proof that propagating plasmons retained the quantum coherency of the photons that launched them. 

In this work, we demonstrate that nonlocal interactions of entangled photons can be used to achieve nonlocal control of light's absorption through the excitation of plasmonic modes. In order to demonstrate the nonlocal control of dissipation with polarization-entangled photons, we constructed a polarization-sensitive 'quantum eraser' interferometer for which we show that the conditions for interference can be non-locally controlled through a polarization-sensitive detection of the entangled photons. We previously demonstrated that the level of dissipation of a thin absorber placed in a single photon interferometer can be varied from nearly $0\%$ ('perfect transmission') to $100\%$ ('perfect absorption'), depending on the position in the standing wave, whereas only $50\%$ absorption is observed, when the standing wave is not formed \cite{16}. Therefore, by enabling interference and controlling the resulting standing wave, one can nonlocally select  the level of dissipation of photons of certain polarization states in the absorber. 

\begin{figure*}
 	\centering 
 	\includegraphics[width=16cm]{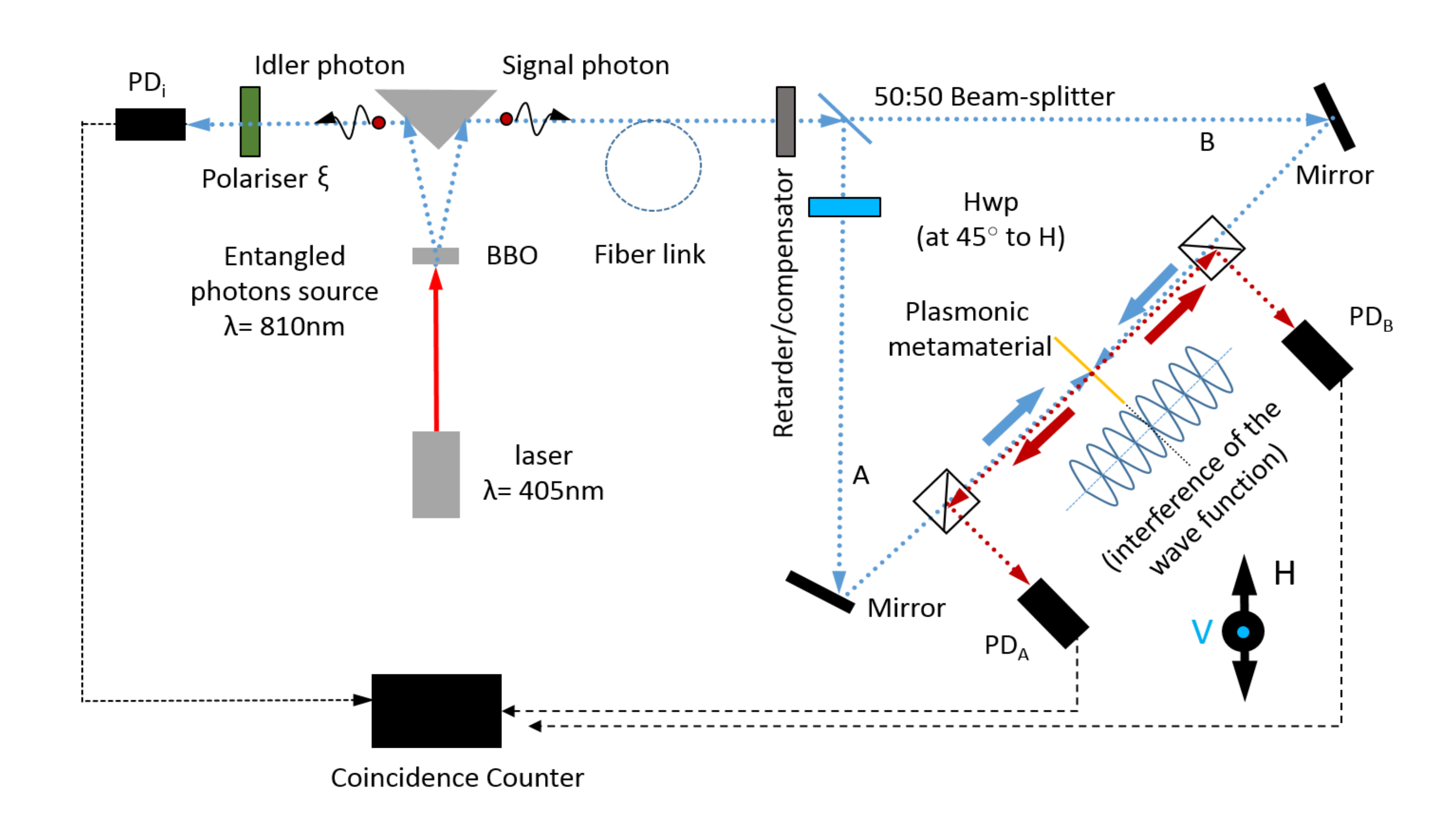}
 	\caption{ Nonlocal dissipation management with entangled photons. Polarization-entangled photon pairs are generated by a type II spontaneous parametric down conversion process (laser at 405 nm impinging a BBO nonlinear crystal). One photon of the pair is introduced in the interferometer (the “signal” photon at the 50/50 beamsplitter) and can take two paths (A and B) with a fiber link and a compensator for synchronization issues. Hwp is a half waveplate in arm A. A thin metamaterial plasmonic absorber is introduced in the polarization-sensitive interferometer. By detecting the idler photon outside the interferometer with the help of a polarizer ($\xi$), one can non-locally prevent or allow deterministic absorption of the signal entangled photon that dissipates through coupling into a plasmon of the absorber placed in the interferometer. $PD_A$, $PD_B$ and $PD_i$ are photon detectors. \label{F:setup}}
 \end{figure*}

{\textit{The concept.---}} In a simplified Sagnac interferometer, an input beamsplitter creates two optical paths A and B (see Fig.\ref{F:setup}). To render the device polarization-sensitive we introduce a half-wave plate only in path A and orient its main axis to be $45^{\circ}$ to the plane of the interferometer \cite{17}. In what follows we will refer to light linearly polarized in the plane of the interferometer as horizontally polarized (H-polarized) light and light polarized perpendicular to the plane as vertically polarized (V-polarized) light. Such an interferometer creates standing waves only for light polarized along the fast and slow axis of the waveplate. Input linear polarizations of $+45^{\circ}$ or $-45^{\circ}$ to the plane of the interferometer will not be affected by the wave-plate and will evolve through both paths A and B as identically polarized traveling waves forming two standing waves in the interferometer with the antinode for $+45^{\circ}$ corresponding to the node for $-45^{\circ}$. Conversely, the half-wave plate converts a vertical polarization into a horizontal one, and vice versa: so if vertically or horizontally polarized light is launched in the interferometer, optical paths A and B will contain counter-propagating orthogonal that do not interfere: a standing wave is not formed in the interferometer. 

Now, with polarization-entangled photon pairs (denoted idler $i$ and signal $s$ photons for each pair, see Fig.\ref{F:setup}) it is possible to nonlocally control the state of the signal photons inside the interferometer, through a measurement on the idler photons outside (and that never enter) the interferometer. This is achieved by adding a polarizer on the idler channel. When the idler polarizer is set to either an angle of $+45^{\circ}$ or $-45^{\circ}$ to the plane of the interferometer, the polarization state of the polarization-entangled signal photon will be polarized at $-45^{\circ}$ or $+45^{\circ}$ correspondingly. The signal photon’s path-entangled wavefunction will form a standing wave in the interferometer and strong dissipation in the 'coherent absorption' regime and zero dissipation in the 'coherent transmission' regime can be observed. Conversely, when the polarizer in the idler channel is set vertically or horizontally, the polarization state of the signal photon will be necessarily projected to the horizontal or vertical polarization, correspondingly. Therefore, due to distinguishability of optical paths \cite{7,8}, no standing wave will be formed in the interferometer and coherent control of the absorption process is removed.

Therefore, the scheme described here is a dissipative form of a 'quantum eraser'. The idler polarizer installed at $+45^{\circ}$ or $-45^{\circ}$ to the plane of the interferometer erases the "which-path” information to the absorber and thus restores the standing wave in the interferometer and thus the coherent absorption regime.

{\textit{The experiment.---}} The experimental setup is shown in Fig.\ref{F:setup}. We generated pairs of polarization-entangled photons at the wavelength of 810 nm by spontaneous parametric down-conversion (SPDC). A 200-mW laser diode with emission centered at the wavelength of $\lambda_p$ = 405 nm was used to pump a 2mm-thick type-II beta-barium borate (BBO) nonlinear crystal producing non-collinear, degenerate photon pairs. Polarization entanglement is achieved by adding 1mm-thick BBO compensation crystals and half wave plates set at $45^{\circ}$. The photon pairs collected from the areas of intersections of phase-matching cones were coupled to single-mode fibres with collimation lenses. A 10-nm bandpass filter centered at 810 nm was used to block the pump radiation and select 'twin' SPDC photons.
As detailed in Fig.\ref{F:setup}, the idler channel was connected to a photon counting avalanche photodiode detector and a coincidence counter (IDQuantique ID800). It was used to control the presence of the signal photon within the interferometer. The signal photons were coupled to the interferometer. A variable retarder was used to compensate for the polarization change in the signal fiber. The photons enter the interferometer through a lossless ($50:50$) non-polarizing beam splitter. 
The thin metamaterial absorber was placed at the centre of the interferometer between two x10 microscope objectives producing a spot size of $\approx 10$ $\mu m$ in diameter. The absorber’s position was scanned using a piezoelectrically actuated linear translation stage over a few optical wavelengths. The sum of the photon counts were detected by the two avalanche detectors ($PD_A$ and $PD_B$) in coincidence with the idler photon $PD_i$ within a 10 ns time window. 

The SPDC source creates a quantum superposition of polarized photons of orthogonal basis. The wavefunction’s general form for polarization entangled states of this type is \cite{20}: 

\begin{equation}\label{1}
 |\Psi\rangle=\frac{1}{\sqrt{2}} \left(|H \rangle_i|V\rangle_s-|V\rangle_i|H\rangle_s \right) 
\end{equation}
 where indices $i$ and $s$ denote the idler and signal photons respectively and $|H\rangle $ and $|V\rangle$ denote the horizontal and vertical polarization states respectively.

 The path entanglement wavefunction of a single photon that enables interference has the general form:
 \begin{equation}\label{2}
 |\Psi\rangle=\frac{1}{\sqrt{2}}\left( |1\rangle_A|0\rangle_B-e^{i\phi}|0\rangle_A|1\rangle_B\right)
 \end{equation}

 By integrating the path entanglement wavefunction Eq.\eqref{2} into the polarization entanglement wavefunction Eq.\eqref{1}, in representation of our optical scheme, we obtain:
\begin{equation}\label{3}
\begin{split}
 |\Psi\rangle= \frac{1}{2}\left[|H\rangle_i\left(|H\rangle_A|0\rangle_B-e^{i\phi}|0\rangle_A|V\rangle_B\right)\right. \\
 \left.-|V\rangle_i\left(|V\rangle_A|0\rangle_B-e^{i\phi}|0\rangle_A|H\rangle_B\right)\right]
\end{split}
 \end{equation}

 And by expanding, we arrive at the path entanglement of two polarization wavefunctions:
\begin{equation}\label{4}
\begin{split}
 |\Psi\rangle=\frac{1}{2}\left[ \left( |H\rangle_i|H\rangle_A-|V\rangle_i|V\rangle_A \right) |0\rangle_B\right.\\
 \left.-e^{i\phi}|0\rangle_A\left( |H\rangle_i|V\rangle_B-|V\rangle_i|H\rangle_B\right)\right] 
 \end{split}
 \end{equation}

We first measured the degree of polarization entanglement of the generated photons . These measurements were performed for two different polarization basis sets, $|H,V\rangle$ and $\pm 45 \rangle $ which correspond to 1) horizontal and vertical polarizations and 2) polarizations at $+45^{\circ}$ and $-45^{\circ}$ to the plane of the interferometer. The Bell parameter was then found to be $S= \sqrt{2} (V_1+V_2)  = 2.66\pm 0.01$ where $V_{1,2}$ were visibilities calculated from the correlation curves for the two basis sets.  Here, according to the Clauser-Horne-Shimony-Holt inequality \cite{20,21,22}, a value of S greater than 2 implies nonlocal quantum correlations. We note that our measured value of the Bell parameter of $S= 2.66$, is close to the maximum value of $S=2\sqrt{2}\approx2.88$ that is expected for perfectly entangled states. 

The plasmonic metamaterial absorber made of split ring resonators was designed to provide a nearly $50\%$ traveling wave absorption, similarly to work reported in Ref.\cite{18}. We fabricated a freestanding gold film of subwavelength thickness nanostructured to create polarization-independent plasmonic absorption at the operational wavelength of 810 nm.

\begin{figure}[t!]
	\centering
	\includegraphics[width=8cm]{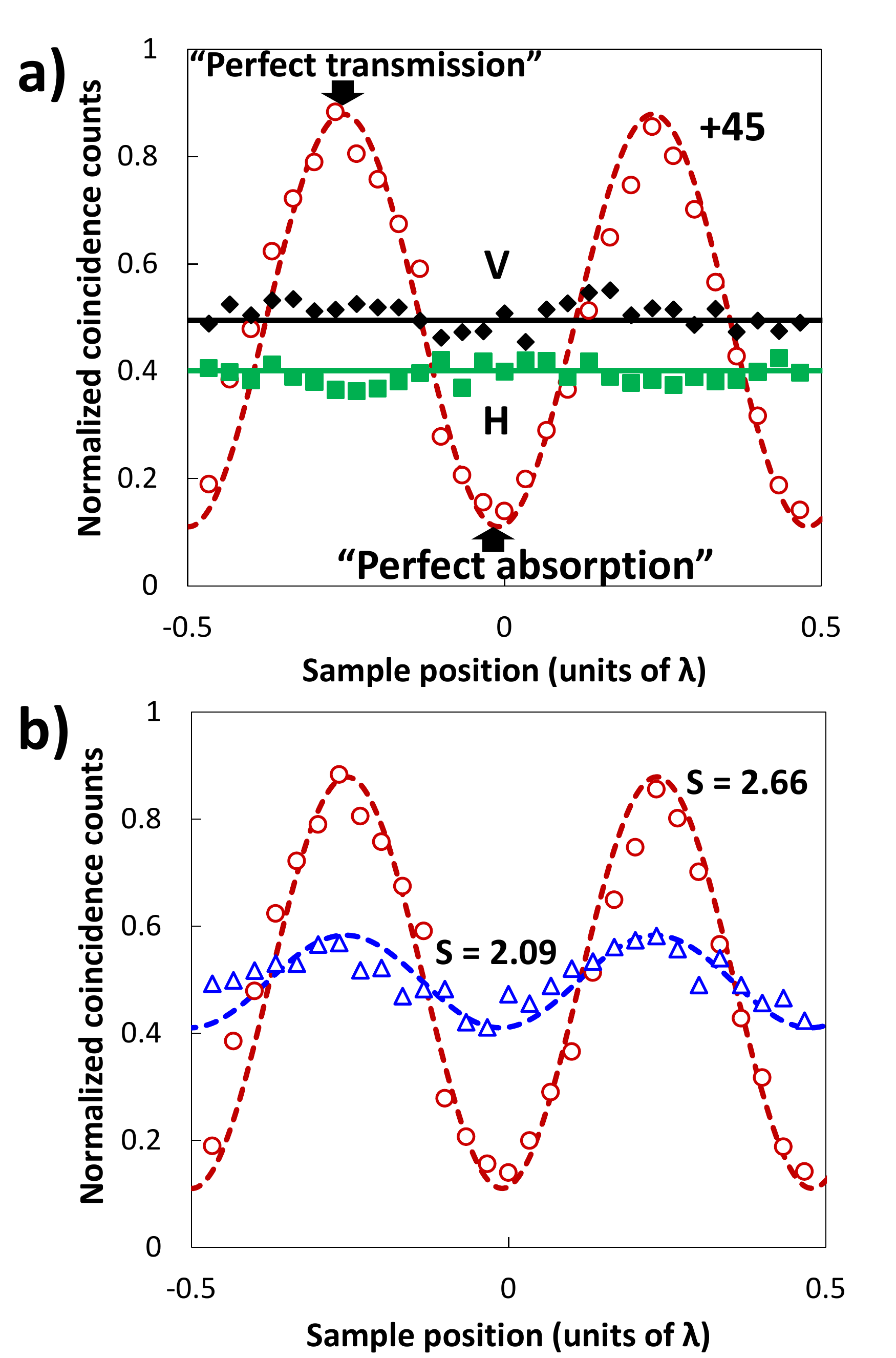}
	\caption{Normalized transmission of the plasmonic thin absorber placed in the interferometer, a) registered for different polarization states of the photons detected in the idler channel with vertical polarization (black diamonds), horizontal polarization (green squares) and with polarization at $45^{\circ}$ to the plane of interferometer (red circles). b) presents the photons detected in the idler channel with polarization at $45^{\circ}$ to the plane of interferometer for two different levels of polarization entanglement of the idler and signal photons: 'strong' entanglement with S=2.66 (red circles) and 'weak' entanglement with S=2.09 (blue triangles).  \label{F:results}}
\end{figure}

To demonstrate the nonlocal control of dissipation with polarization-entangled photons, we introduce one photon of the entangled pair (signal) in the polarization sensitive interferometer where it interacts with the plasmonic metamaterial absorber. We placed the absorber on a piezo-driven actuator in the center of the interferometer (see Fig.\ref{F:setup}). We then detected the level of light intensity at the interferometer output by taking the sum of the photon counts registered by photodetectors $PD_A$ and $PD_B$ that are heralded by the detection events of the idler photons on $PD_i$. We then normalized these to the total level of photon counts when the absorber is removed from the interferometer and recorded the normalized level as a function of absorber’s position along the standing wave. The results of these measurements are presented in Fig.\ref{F:results}.

No dependence of photon counts jointly registered by photodetectors $PD_A$ and $PD_B$ on the position of the plasmonic metamaterial were seen if the heralding was performed with an idler polarization set (with polarizer $\xi$ in Fig.\ref{F:setup}) to vertical or horizontal (curves with  black diamonds for V and green squares for H in Fig.\ref{F:results}-a). This measurement and distinguishability of which-path information derived from Eq.\eqref{4} can be described by:
\begin{equation}\label{5}
\langle H|\Psi\rangle=\frac{1}{2} \left(|H \rangle_A|0\rangle_B-e^{i\phi}|0\rangle_A|V\rangle_B \right) 
\end{equation}
Thus, the single photon wavefunction does not interfere or form a standing wave in the interferometer: coherent control of absorption is lost and photons entering the metamaterial film suffer probabilistic absorption of approximately $50\%$.  A small difference in the level of absorption between curves for vertical and horizontal polarizations is explainable by residual anisotropy of the plasmonic absorber.

On the contrary, if heralding is performed with the idler polarization set to $+45^{\circ}$ with respect to the plane of the interferometer, we observe a clear oscillation of absorption as a function of the absorber's position in the standing wave (curve with red circles in Fig.\ref{F:results}-a). The which-path information has been erased and the path-entangled single photon wavefunction forms a standing wave in the interferometer: 
\begin{equation}\label{6}
\langle +45|\Psi\rangle=\frac{1}{2} \left(|-45 \rangle_A|0\rangle_B-e^{i\phi}|0\rangle_A|-45\rangle_B \right) 
\end{equation}
The corresponding modulation of the absorption/transmission is  shown in Fig.\ref{F:results}-a.  We note that when absorption is achieved for $+45^{\circ}$, transmission is correspondingly observed for $-45^{\circ}$. Each entangled  photon entering the interferometer is therefore deterministically absorbed and converted into a plasmon in the nanostructure. 

Finally, in order to underline the role of polarization entanglement, we compared the results from experiments performed with two levels of polarization entanglement. We adjusted our photon source from the regime when it generated entangled states close to maximum Bell parameter $S=2.657\pm 0.004$, to a rather low degree of entanglement with $S=2.087 \pm 0.004$. The source of photons with a lower degree of entanglement yielded absorption modulation of approximately $14\%$ (see Fig.\ref{F:results}-b, curve with blue triangles) compared to the $80\%$ (see Fig.\ref{F:results}-b, curve with red circles) attainable with strongly entangled photons. 

The reduction of the absorption modulation visibility in the experiments performed with photon pairs with high and low values for the Bell parameter is relevant to the fundamental differences in polarization properties of the generated photons. The regime of high Bell parameter implies that photons traveling in the same direction, but emerging from the intersections of two phase-matching cones of the parametric down conversion crystal can create arbitrary superpositions of vertically and horizontally polarized states defined by the nonlinear down-conversion process. As a result entangled, orthogonally polarized pairs of photons are generated with arbitrary polarization basis. In contrast, in the regime of low values of the Bell parameter, quantum superpositions of vertically and horizontally polarized states are not formed and the parametric device generates orthogonally polarized photons where idler and signal photons can either be horizontally or vertically polarized.  This mixed state produced is thus not sufficient to produce interferences, unlike an entangled state.

{\textit{Conclusion.---}}
In conclusion, we have demonstrated a regime of nonlocal control of dissipation of polarization-heralded photons. By selecting the idler photon polarization that never entered the interferometer, we can switch the metamaterial from the regime of travelling-wave absorption to the regime of coherent absorption. We therefore demonstrated a new type of quantum gate for which the output signal can be switched nonlocally in the 'coherent transmission' regime from unitary transmission to a probabilistic $50\%$ transmission. Alternatively, in the 'coherent absorption' configuration, the system can be switched nonlocally from zero transmission to a probabilistic $50\%$ transmission. 

It should be noted that control of dissipation of polarization-heralded photons does not imply that the total dissipation of light energy on the absorber can be nonlocally controlled. Indeed, photons with other polarization states are always simultaneously present and also enter the energy balance making the total dissipation independent from the heralding or nonlocal control. 

The  approach shown here can be used not only for nonlocal control of coupling of photons to localized plasmons, but can also be exploited for the nonlocal control of coupling of photons to plasmon polaritons in polarization-sensitive schemes where photon-plasmon entanglement can be envisaged.

Following a period of embargo, the data from this article can be obtained from the University of Southampton ePrints research repository,
http://dx.doi.org/10.5258/SOTON/xxxxx

{\textit{Acknowledgements.}} The authors acknowledge the support of the Singapore MOE Grant MOE2011-T3-1-005, EPSRC (U.K.) grants EP/M009122/1 and EP/J00443X/1 and EU Grant ERC GA 306559.  C.C. would like to thank the Champagne-Ardenne region, the French LABEX Action and the EU COST Action Nanoscale Quantum Optics. Authors acknowledge N. Zheludev for inspiring and supporting this work and T. Rogers and G. Adamo for useful discussions.

\end{document}